\documentclass[a4paper,11pt]{article}
\usepackage{pos}

\usepackage{amsmath}
\usepackage{amssymb}
\usepackage{caption}
\usepackage{subcaption}
\usepackage{csvsimple}
\usepackage{multirow}
\usepackage{booktabs}

\usepackage{lineno}

\title{Searching for High-Energy Neutrinos from Core-Collapse Supernovae with IceCube}

\ShortTitle{High-Energy Neutrinos from Core-Collapse Supernovae}

\author{The IceCube Collaboration \\{\normalsize \normalfont(a complete list of authors can be found at the end of the proceedings)}}

\emailAdd{jannis.necker@desy.de}

\abstract{

IceCube is a cubic kilometer neutrino detector array in the Antarctic ice that was designed to search for astrophysical, high-energy neutrinos. It has detected a diffuse flux of astrophysical neutrinos that appears to be of extragalactic origin. A possible contribution to this diffuse flux could stem from core-collapse supernovae. The high-energy neutrinos could either come from the interaction of the ejecta with a dense circumstellar medium or a jet, emanating from the star's core, that stalls in the star's envelope. Here, we will present results of a stacking analysis to search for this high-energy neutrino emission from core-collapse supernovae using 7 years of $\nu_\mu$ track events from IceCube.\\

\vspace{4mm}
{\bfseries Corresponding authors:}
Jannis Necker$^{1*}$\\
{$^{1}$ \itshape DESY}\\[4mm]
$^*$ Presenter

\FullConference{37$^{\rm{th}}$ International Cosmic Ray Conference (ICRC 2021)\\
		July 12th -- 23rd, 2021\\
		Online -- Berlin, Germany}

}

\begin{document}
\maketitle


\section{Introduction}\label{sec:info}

IceCube, completed in 2010, is a cubic-kilometer neutrino detector installed in the ice at the geographic South Pole \cite{Aartsen:2016nxy} between depths of 1450 m and 2450 m. Reconstruction of the direction, energy and flavor of the neutrinos relies on the optical detection of Cherenkov radiation emitted by charged particles produced in the interactions of neutrinos in the surrounding ice or the nearby bedrock. The IceCube Collaboration has measured an astrophysical flux of neutrinos, consistent with a simple power law. This flux has been characterised with different detection channels such as up-going muon tracks \cite{Aartsen2016xlq}, high-energy starting events \cite{Abbasi2020jmh} and cascade events \cite{Aartsen2020aqd}. The result of a joint fit of these channels is consistent with a spectral index of $\gamma = -2.5$ \cite{diffuse_flux_measurement}. Despite extensive searches, the sources of this flux still remain largely unknown.

Core-Collapse Supernovae (SNe) can potentially produce high-energy neutrinos. When the exploding star is surrounded by a dense circumstellar medium (CSM), shocks can be induced by the SN ejecta that enable particle acceleration \cite{MuraseThompson2011}. Signs of this interaction can be seen in the spectra of SNe IIn and also IIP. The progenitors of SNe Ibc are believed to be massive Wolf-Rayet stars that typically develop heavy winds at the end of their lives which could provide a sufficiently dense CSM.
Furthermore, a jet can emanate from the collapsing core of the star and particle acceleration can occur in internal shocks and at the jet head. If this jet is not energetic enough to break through the remaining outer layers of the star and stalls sufficiently far below the photosphere, the high-energy neutrinos produced by accelerated protons would be the only messengers to escape \cite{SennoMurase2016}. The massive progenitors of SNe Ibc are good candidates for this scenario.

Here, we describe the results for a search for high-energy neutrinos from optically-detected core-collapse supernovae. The search used 7 years of data from IceCube collected between April 2008 and May 2015. The data sample contains 711,878 through-going track events with a mean angular resolution of about $1^\circ$ created by $\nu_\mu$ interactions in the ice and bedrock. The reconstructed energies of the neutrino events range between 7 GeV and 57 PeV with a median energy of 2 TeV.


\section{Emission Models and Catalogues}\label{sec:catalogue}

The high-energy neutrino emission in the CSM scenario will take place on a timescale from months up to years after the explosion. Cosmic rays are accelerated by shocks in the CSM induced by the supernova ejecta and will produce high-energy neutrinos via inelastic proton-proton scattering. This produces mesons that decay into gamma-rays and TeV neutrinos. In the following we describe two time evolution models that we use in the analysis.

Assuming that $\sim$ 10\% of the supernova explosion energy goes into the acceleration of cosmic rays, the model presented in \cite{MuraseThompson2011} suggests the emission of the high-energy neutrinos lasts until about 100 to 1000 days after the explosion while being independent of the evolution of the emission strength within that time. Following this model, we adopt box time models with lengths of 100, 300 and 1000 days. We use these models for all supernova types.

The second time evolution model is based on considerations presented in \cite{ZirakashviliPtuskin2016}. Assuming the density profile of the CSM decreases with distance $r$ from the explosion as $\rho\sim r^{-2}$, and modeling the production of cosmic rays via non-linear diffusive shock acceleration, the high-energy neutrino flux is $\Phi(t)$ is predicted to decrease from an initial flux $\Phi_0$ according to 
\begin{equation}
    \Phi(t) = \Phi_0 \left(1+ \frac{t}{t_\mathrm{pp}} \right)^{-1}
\end{equation}
where $t_\mathrm{pp}$ is the proton-proton interaction timescale.
As $t_\mathrm{pp}$ is responsible for the fading of the neutrino emission we refer to it as the decay time. For typical explosion parameters of SNe IIn it is about $0.2 \, \mathrm{years}$. To also test slightly shorter and longer timescales we adopt the decay model with decay times of 0.02, 0.2 and 2.0 years. We apply this second model to SNe IIn and IIP only.

Finally, in the choked-jet scenario that can occur in SNe Ibc, the neutrino emission is expected in a short time window ($\sim$ 1 day) around the explosion. 
Because the precise explosion time of the SNe Ibc is not known, we search using a box model spanning 20 days prior to the first detection of the explosion. 
Table \ref{tab:pvalues} indicates the flux models applied to the various supernova types.

The SN catalogues were assembled from the \textit{WiseRep} catalogue \cite{WISErep}, the \textit{ASAS-SN} survey data \cite{HoloienStanek2016} and the \textit{Open Supernova Catalogue} \cite{GuillochonParrent2017}. When a SN is discovered by different observers they sometimes assign different event names and an association is not always performed. The catalogue entries were therefore merged, accounting for these different aliases of the same events by requiring a separation in time by at least 50 days for events that have an angular separation of less than $0.1^\circ$. The resulting catalogues contain 387 IIn, 167 IIP and 824 Ibc SNe.

Because the method outlined in section \ref{sec:methods} gets computationally expensive for a large number of sources, we define a high-quality subsample of nearby sources. Assuming an $E^{-2}$ neutrino spectrum, we calculate the expected flux of each source. We do not make detailed assumptions about individual sources and assume the same intrinsic flux. Under these assumptions, the high-quality subsample is defined as the brightest sources that contribute 70\% of the total neutrino flux from the respective supernova type. This final sample then contains 15 SNe IIn, 20 SNe IIP and 19 SNe Ibc.


\section{Analysis Method}\label{sec:methods}

\subsection{Maximum Likelihood Method}

We perform an unbinned maximum likelihood analysis to look for a correlation between the supernova catalogues and the neutrino data using the point source search outlined in \cite{txs_analysis}. The likelihood is described as a linear combination of the signal PDF $\mathcal{S}$ and background PDF $\mathcal{B}$, that depend on the neutrino properties $\theta$. The PDFs are evaluated for each event $i$ and the likelihood is constructed as a product of the independent $N$ events,
\begin{align}
\label{eq:llh_w_pdfs}
    \mathcal{L}(n_{\mathrm{s}}, \gamma) = \prod _{\mathrm{i=0}} ^\mathrm{N} \left[ \frac{n_{\mathrm{s}}}{\mathrm{N}}\mathcal{S}(\theta_\mathrm{i}, \gamma)  + \left( 1- \frac{n_{\mathrm{s}}}{\mathrm{N}} \right) \mathcal{B}(\theta_\mathrm{i})  \right],
\end{align}
where $\gamma$, the index of the assumed power law spectrum, and $n_\mathrm{s}$, the number of signal neutrinos, are free parameters.
Both PDFs can be written as a product of their energy, spatial and temporal components. 
We assume the background, which is dominated by atmospheric neutrinos, to be uniform over the livetime $\tau$ and model the spatial and energy part from the dataset, assuming it is dominated by background. The background PDF then only depends on the declination $\delta$ and the energy $E$:
\begin{align}
    \mathcal{B}(\delta, E) = \frac{1}{2\pi} f_\delta(\delta) \cdot \frac{1}{\tau} \cdot f_\mathrm{E}(E, \delta)
\end{align}

For the spatial component of the signal PDF, we assume a two-dimensional circular Gaussian distribution centered on the source position and with a width equal to the angular uncertainty $\sigma$ of each neutrino event's reconstructed direction. It depends on the angular separation of the a source position $\Vec{s}$ and an event position $\Vec{x}(\delta, \alpha)$ that is given by $r^2(\delta, \alpha) = (\Vec{s} - \Vec{x}(\alpha, \delta))^2$.

The signal energy distribution as a function of direction is obtained using Monte-Carlo simulations, which typically assume an unbroken power law $E^{-\gamma}$. Hard spectra, e.g. $\gamma = 1$, produce more high-energy neutrinos that are easily distinguishable from atmospheric background events whereas for soft spectra, e.g. $\gamma = 4$, the distribution gets more similar to the background distribution. The signal energy PDF $f_\mathrm{E}$ is therefore dependent on the declination and energy, as well as the assumed spectral index.

As mentioned in section \ref{sec:catalogue}, we assume several models for the time-dependent  neutrino number luminosity $L_\mathrm{\nu}(t)$. We construct a temporal PDF by dividing $L_\mathrm{\nu}(t)$ by its time integral over the range $t_\mathrm{start}$ to $t_\mathrm{end}$, assuming $L_\mathrm{\nu}(t)$ is zero outside of this range.
For a box-shaped lightcurve the limits of integration are simply its start and end time. For the decay model we take $t_\mathrm{end}$ to be the time when the flux decreases to $0.1\%$ with respect to the initial flux.

Combining the spatial, energy and temporal PDFs, the signal PDF can then be written as 
\begin{equation}
    \mathcal{S}(\alpha, \delta, \sigma, E, \gamma, t) = \frac{1}{2\pi \sigma^2} \, e^{\frac{-r^2(\alpha, \delta)}{2\sigma^2}} \cdot f_\mathrm{E}(E, \delta, \gamma) \cdot \frac{\mathrm{L}_\mathrm{\nu}(t)}{\int_{\mathrm{t_{start}}} ^{\mathrm{t_{end}}} \, \mathrm{L}_\mathrm{\nu}(t') \, \mathrm{d}t'}
\end{equation}
When stacking several sources, the signal PDF becomes a weighted sum $\mathcal{S} =  \sum_j \, \omega_j \mathcal{S}_j$ over the PDFs describing each individual source $j$. The weights can be estimated using prior knowledge about the sources from other measurements, such as their bolometric luminosity. The point source stacking likelihood then becomes

\begin{equation}
\label{eq:stack_llh}
    \mathcal{L}(n_{\mathrm{s}}, \gamma) = \prod _{\mathrm{i=0}} ^\mathrm{N} \left[ \frac{n_s}{\mathrm{N}} \sum_\mathrm{j=0} ^\mathrm{M} \omega_\mathrm{j} \cdot \mathcal{S}_\mathrm{j}(\theta_\mathrm{i}, \gamma)  + \left( 1- \frac{n_{\mathrm{s}}}{\mathrm{N}} \right) \mathcal{B}(\theta_\mathrm{i})  \right]
\end{equation}
To compare the signal hypothesis to the background hypothesis we maximize the likelihood ratio test statistic
\begin{equation}
    \lambda = 2 \cdot \log \left( \frac{\mathcal{L}(n_\mathrm{s}, \gamma)}{\mathcal{L}(0)} \right)
\end{equation}

\subsection{Fitting the weights}

In this analysis we do not have a way of estimating the relative source contributions and thus determining the source weights $\omega_\mathrm{j}$. Sub-optimal weighting will decrease our sensitivity since weak sources that get a high weight will only contribute background, while the signal of strong sources that get a small weight will be suppressed. 
To circumvent this we choose a model-independent approach and include independent contributions from each source  $\vec{n_\mathrm{s}} = (n_\mathrm{s, 1}, n_\mathrm{s, 2}, ..., n_\mathrm{s, M})$, where $n_s=\sum_j n_{s,\mathrm{j}}$, as free parameters in the fit. As this increases the number of free parameters to $M + 1$ for a catalogue with $M$ sources, the maximization of the likelihood is computationally expensive and can only be performed for a catalogue of limited size as outlined in section \ref{sec:catalogue}. 
The final likelihood is then
\begin{equation}
\label{eq:stack_llh_fit}
    \mathcal{L}(\vec{n_{\mathrm{s}}}, \gamma) = \prod _{\mathrm{i=0}} ^\mathrm{N} \left[ \frac{1}{\mathrm{N}} \sum_\mathrm{j=0} ^\mathrm{M} n_\mathrm{s,j} \cdot \mathcal{S}_\mathrm{j}(\theta_\mathrm{i}, \gamma)  + \left( 1- \frac{n_{\mathrm{s}}}{\mathrm{N}} \right) \mathcal{B}(\theta_\mathrm{i})  \right]
\end{equation}
This is the likelihood function we use in this analysis.


\section{Results}\label{sec:results}

\subsection{Significance Testing}
Using seven years of IceCube muon track events from April 2008 until May 2015 \cite{IceCube7yrdata}, we maximize the likelihood for every scenario in Table \ref{tab:pvalues} and compare the resulting best fit test statistic with a distribution of test statistic values obtained from $2\times10^5$ background-only trials. The resulting p-values are listed in Table \ref{tab:pvalues}. Because we do not allow negative values for the test statistic, we quote a p-value of $>0.50$ where we find a best fit test statistic smaller than the background median. 
We find the most significant results are $p=0.09$, corresponding to the box model fit to SNe IIn with a length of 100 days, and $p=0.06$ for the 300-day box model. Neither result is statistically significant.

\begin{table}[]
    \centering
    \begin{tabular}{c cccc ccc}
        \toprule
        \multirow{2}{*}{ Catalogue} & \multicolumn{4}{c}{Box models [d]} & \multicolumn{3}{c}{Decay models [y]} \\
        & 20 & 100 & 300 & 1000 & 0.02 & 0.2 & 2.0 \\
        \midrule 
        & \multicolumn{7}{c}{p-values} \\
        IIn & - & 0.09 & 0.06 & >0.5 & 0.19 & 0.26 & >0.5 \\
        IIP & - & 0.49 & >0.50 & 0.28 & >0.50 & >0.50 & 0.30 \\
        Ibc & >0.50 & >0.50 & >0.50  & 0.35 & - & - & - \\
        \bottomrule
    \end{tabular}
    \caption{The pre-trial p-values for the correlation of 7 years of IceCube data and the selected supernova catalogues. We quote a p-value of $>0.50$ where the measured test statistic is above the background median. We observe mildly significant over-fluctuations for SNe IIn for the agnostic box models with lengths 100 and 300 days respectively.}
    \label{tab:pvalues}
\end{table}

\begin{figure}
    \centering
    \begin{subfigure}[b]{0.48\textwidth}
        \centering
        \includegraphics[width=\textwidth]{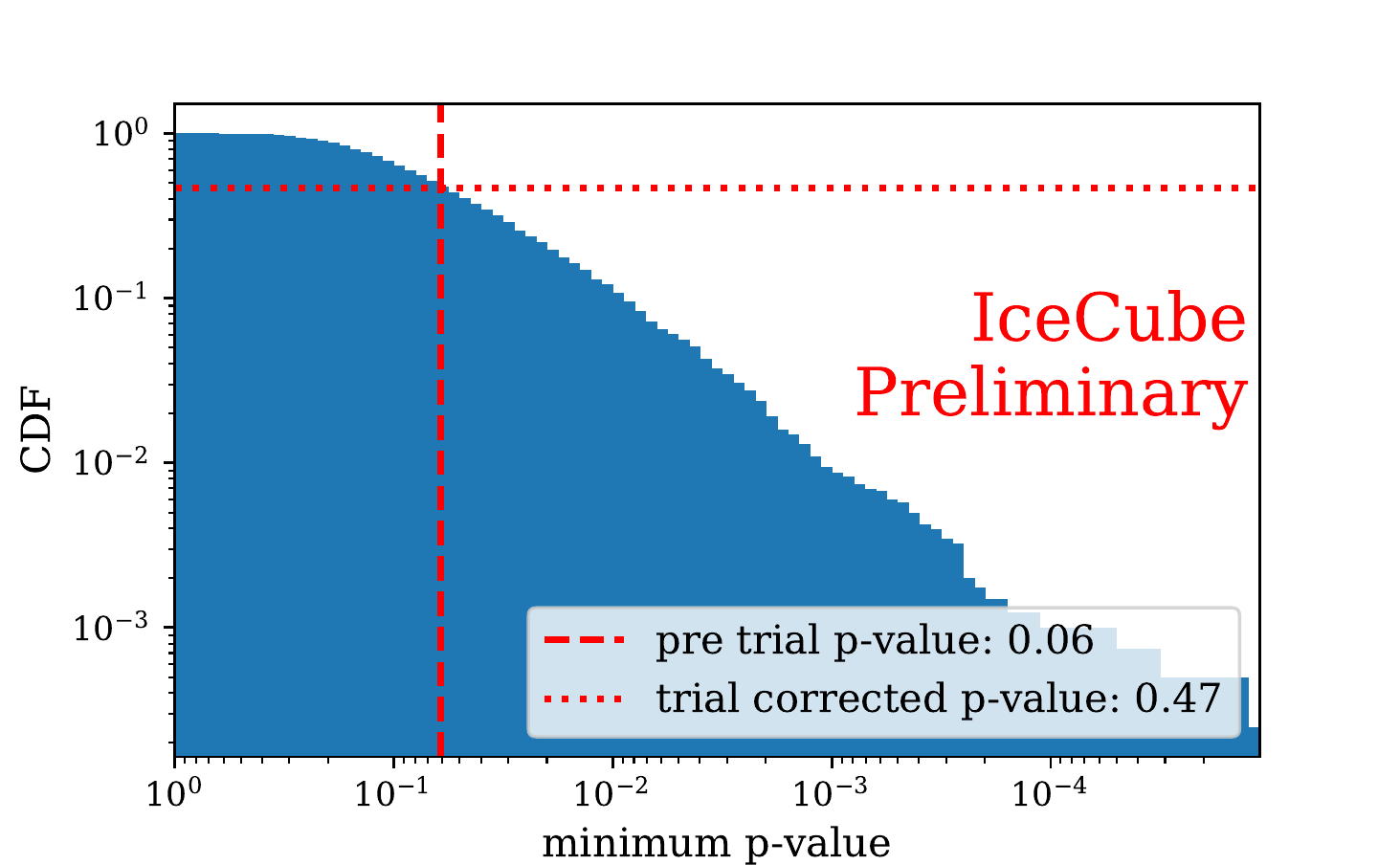}
        \caption{Trial correction of the minimum p-value.}
        \label{subfig:min_pval_correct}
    \end{subfigure}
    \begin{subfigure}[b]{0.48\textwidth}
        \centering
        \includegraphics[width=\textwidth]{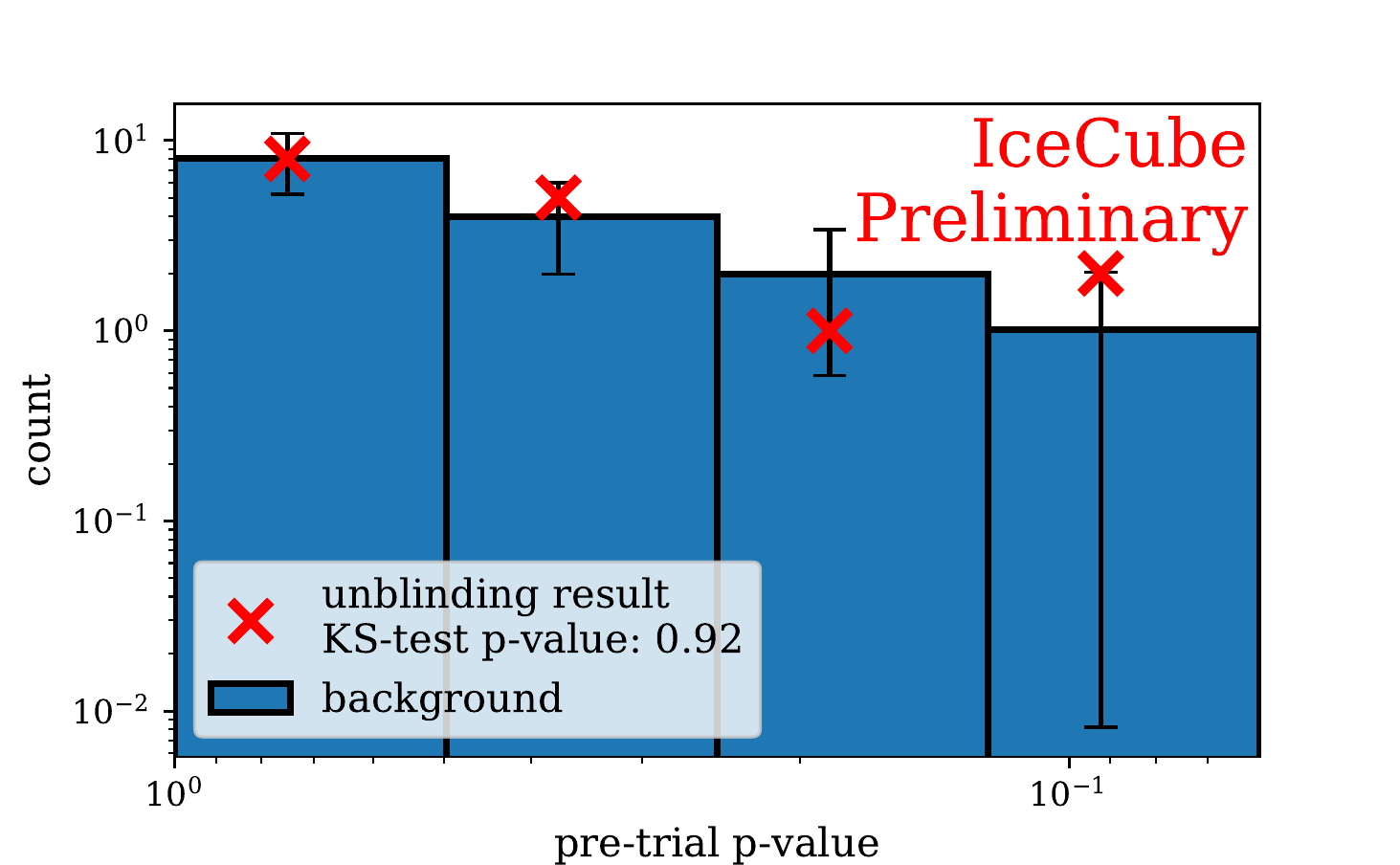}
        \caption{P-value distribution compared to background.}
        \label{subfig:pval_distr_correct}
    \end{subfigure}
    \caption{Visualisation of the trial correction of the p-values. We compare the minimum pre-trial p-value to a distribution of minimum p-values from background and the full distribution of p-values to a background distribution of p-values.}
    \label{fig:my_label}
\end{figure}

To correct our p-values for statistical trials, we compare the lowest p-value $p_{\mathrm{min}} = 0.06$ to the cumulative distribution of lowest p-values obtained with $4 \times 10^3$ background simulations (see figure \ref{subfig:min_pval_correct}). We find that we can expect a lowest p-value at least as small as 0.06 in about half the cases, so the resulting trial-corrected p-value is $p_{\mathrm{min, corrected}} = 0.47$.

Although the individual p-values are not significant, it would still be possible that the distribution of p-values is inconsistent with background expectations. To evaluate the compatibility, we compare it to the distribution of p-values from $4 \times 10^3$ background trials using a \textit{Kolmogorov-Smirnov-Test} (see figure \ref{subfig:pval_distr_correct}). We obtain a p-value of $p_{\mathrm{KS}} = 0.92$ and conclude that the distribution of p-values is compatible with the distribution expected from background. 

\subsection{Upper Limits on the Neutrino Flux from Supernovae}
\begin{figure}
    \centering
    
    \begin{subfigure}[b]{0.48\textwidth}
        \centering
        \includegraphics[width=\textwidth]{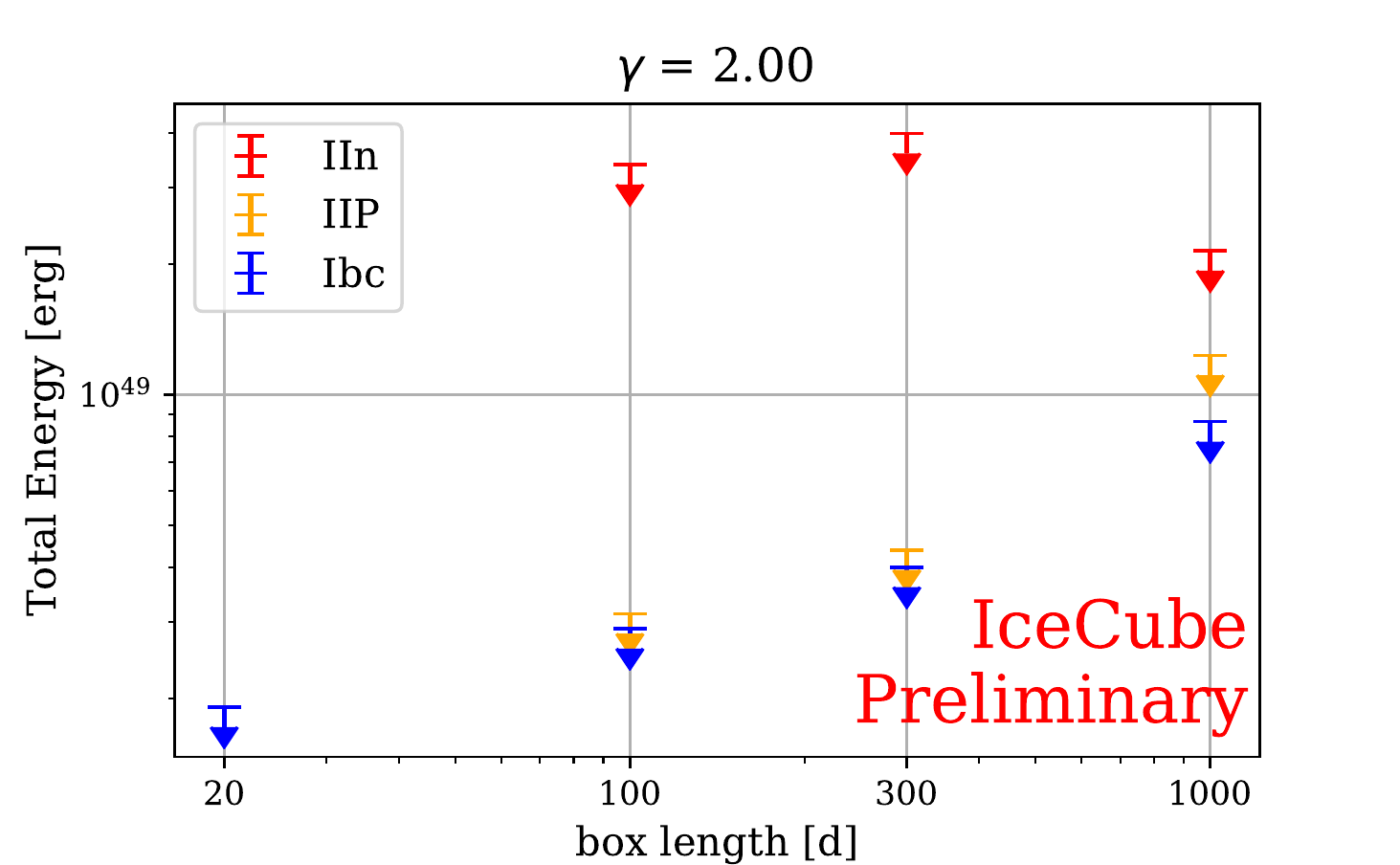}
        \caption{Box time models, $\gamma = 2.0$}
    \end{subfigure}
    \begin{subfigure}[b]{0.48\textwidth}
        \centering
        \includegraphics[width=\textwidth]{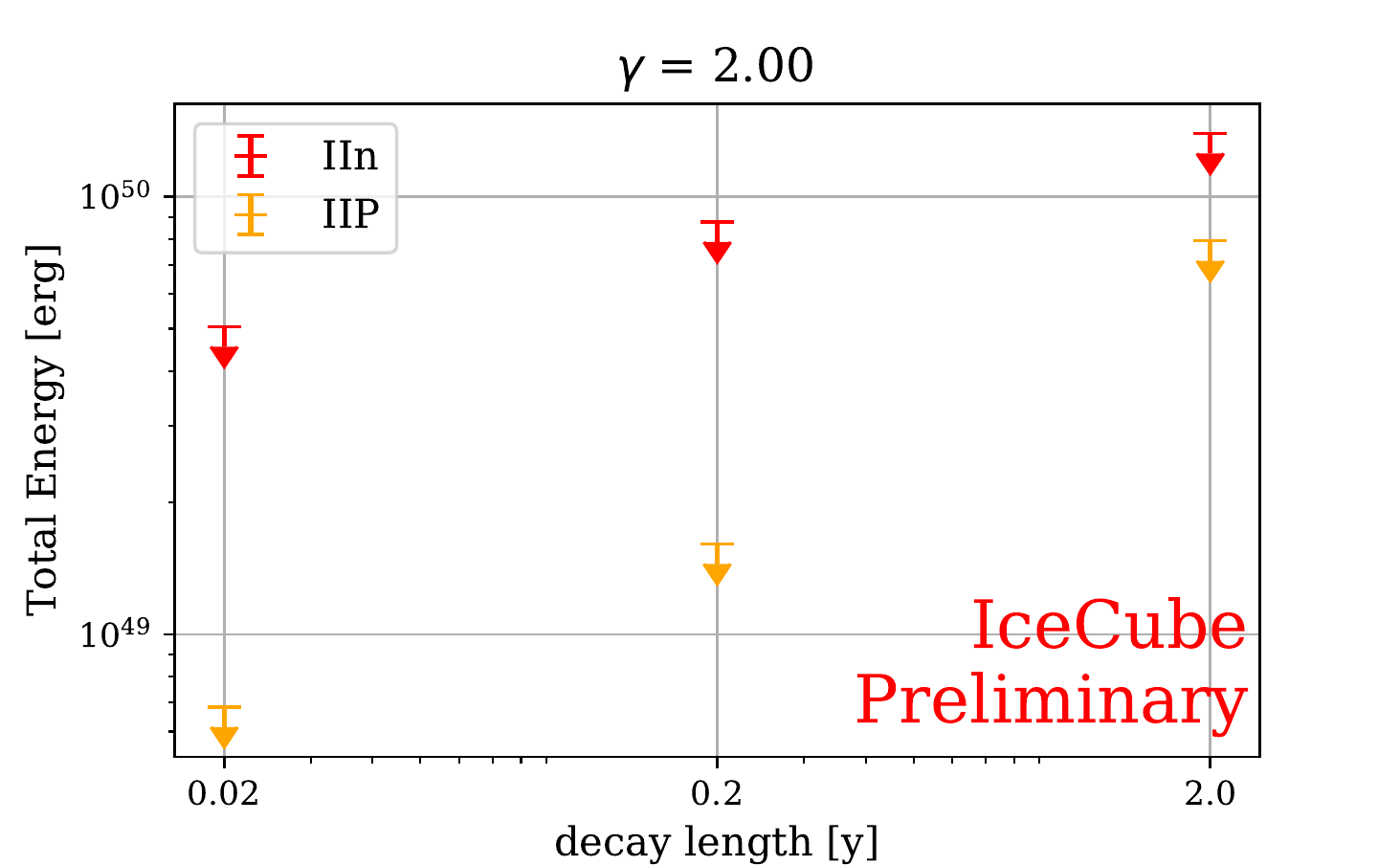}
        \caption{Decay time models, $\gamma = 2.0$}
    \end{subfigure}
    \begin{subfigure}[b]{0.48\textwidth}
        \centering
        \includegraphics[width=\textwidth]{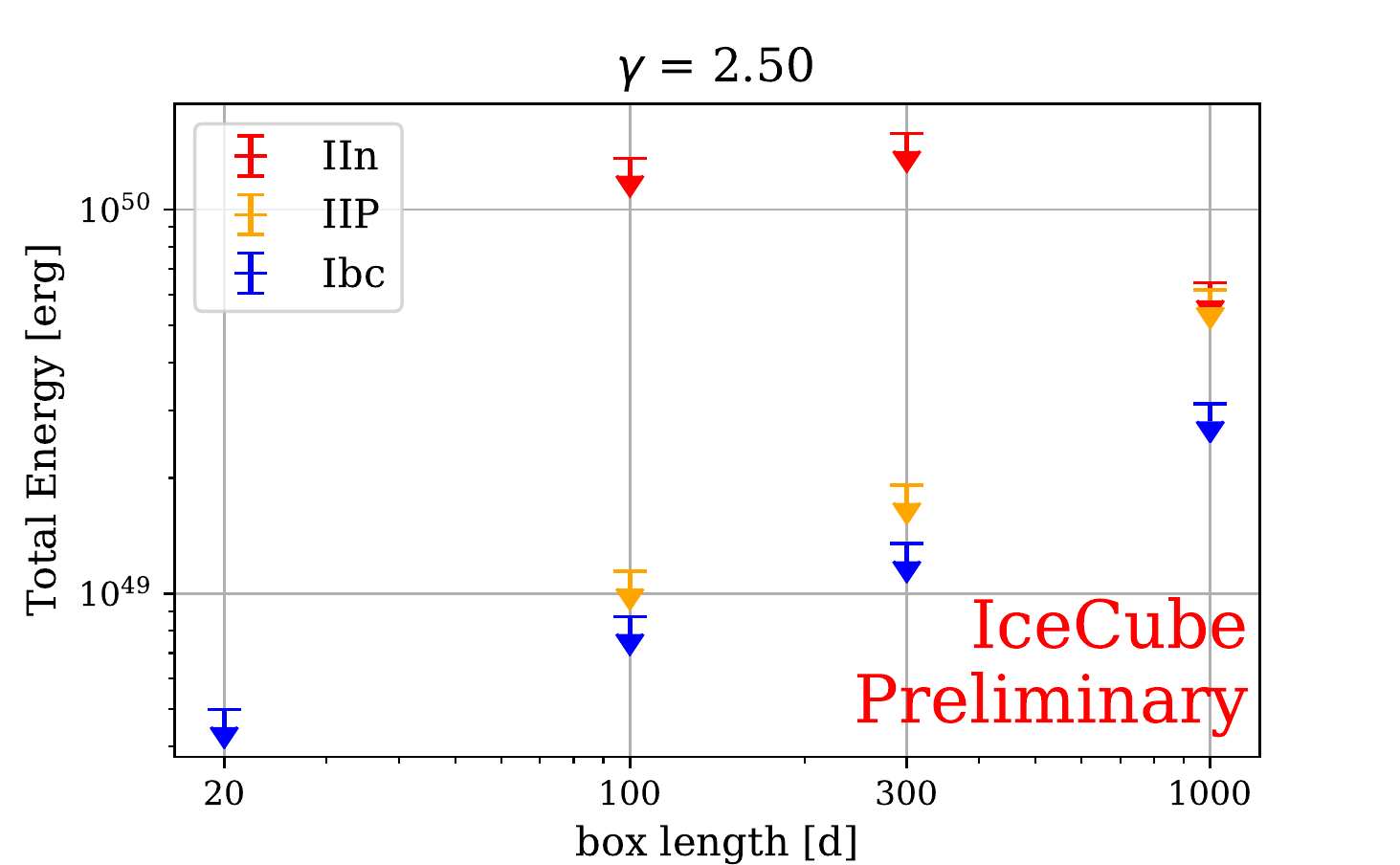}
        \caption{Box time models, $\gamma = 2.5$}
    \end{subfigure}
    \begin{subfigure}[b]{0.48\textwidth}
        \centering
        \includegraphics[width=\textwidth]{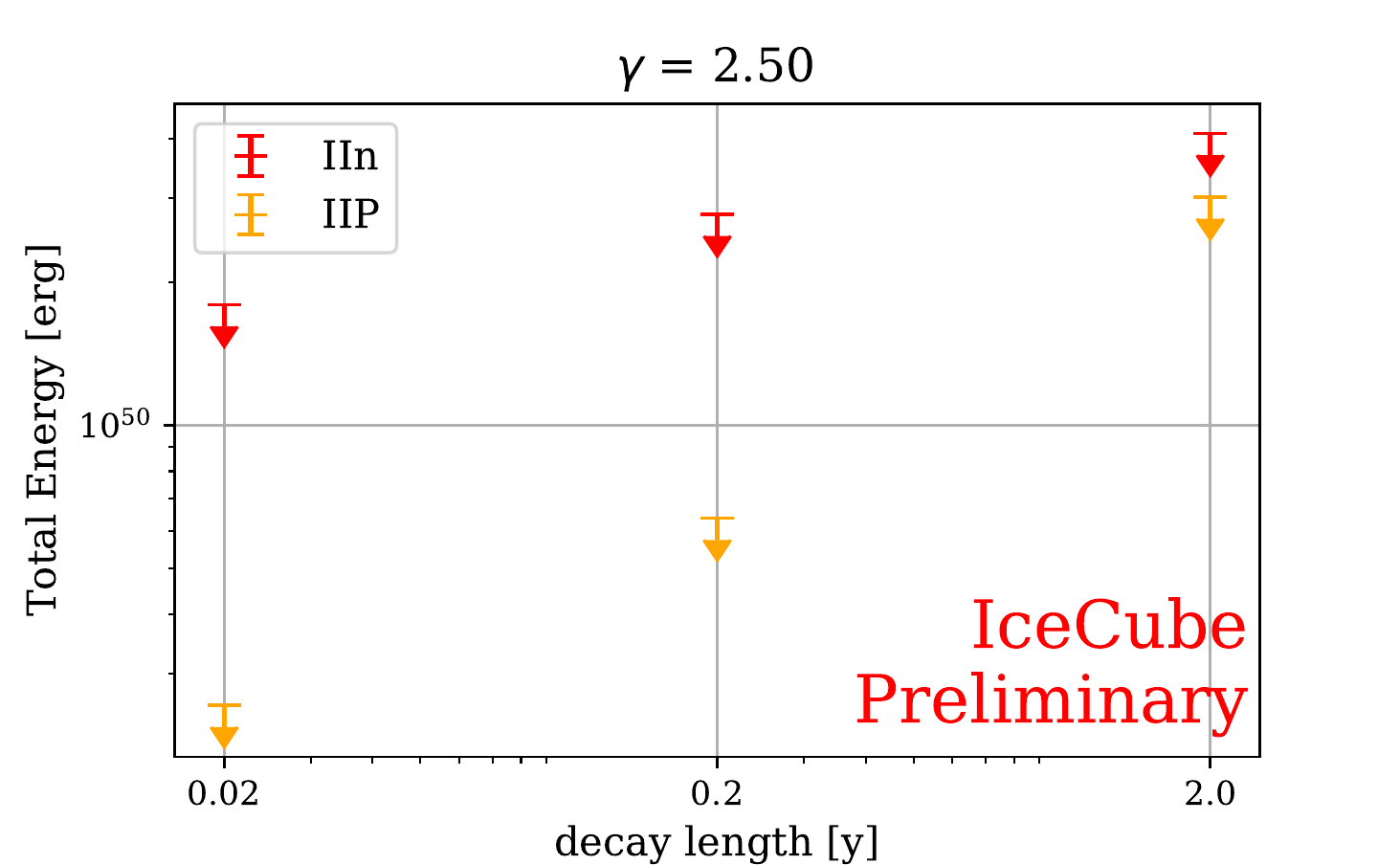}
        \caption{Decay time models, $\gamma = 2.5$}
    \end{subfigure}
    
    \caption{Limits on the total emitted isotropic energy equivalent in neutrinos. Where no overfluctuations are measured, those are derived from the sensitivity. The most constraining case for the CSM interaction scenario for SNe Ibc and IIP is the box model with a length of 100 days and the one with a length of 1000 days for SNe IIn.}
    \label{fig:energy_limits}
\end{figure}

We derive the 90\% upper limit on the flux normalisation from the measured test statistic. Where that limit is more constraining than the one derived from the background median, we use the background median as a floor.
Assuming an unbroken power law spectrum with an index $\gamma$ we can derive limits on the total emitted energy. Further, we assume isotropic emission of each SN and give the results for assumed spectral indices of $\gamma = 2$ and $\gamma = 2.5$, which are shown in figure \ref{fig:energy_limits}. We find the most constraining scenario for SNe IIp and Ibc in the CSM interaction scenarios to be the agnostic box model with a length of 100 days, while the box model with 1000 days length is most constraining for SNe IIn. For a spectral index of $\gamma= 2.5$ we constrain SNe IIn, IIP and Ibc to emit no more than $6.4 \times 10^{49}$,  $1.1 \times 10^{49}$ and $8.7 \times 10^{48}$ erg, respectively, assuming CSM interaction. Assuming the choked-jet scenario SNe Ibc emit no more than $5.0 \times 10^{48}$ erg.

Assuming that our catalogues are representative subsamples of the respective supernova populations, we can calculate limits on their contribution to the diffuse astrophysical neutrino flux measured by IceCube. The flux is given by
\begin{equation} \label{eq:population_flux}
    \Phi(E) = \frac{c}{H_0} \int_0 ^{\infty} \frac{\mathrm{d}N}{\mathrm{d}E} \frac{\rho(z)}{(1+z)} \frac{\mathrm{d}z}{\sqrt{\Omega_{\mathrm{\Lambda}} + \Omega_{\mathrm{k}}(1+z)^2 + \Omega_{\mathrm{m}}(1+z)^3}}
\end{equation}
where $\mathrm{d}N / \mathrm{d}E$ is the neutrino number energy spectrum that we take to follow a power law with spectral index $\gamma = 2.5$ to match the diffuse flux measurement. The cosmological parameters $H_0$, $\Omega_{\mathrm{\Lambda}}$, $\Omega_{\mathrm{k}}$ and $\Omega_{\mathrm{m}}$ are taken from Planck measurements \cite{planck2016}. The volumetric supernova rates $\rho(z)$ are taken from \cite{Strolger_2015} and we use the percentage of supernova subtypes from \cite{Li2011}. Integrating up to a redshift of 2, we find that SNe  IIn, IIP and Ibc can contribute up to 55.2\%, 79.6\% and 28.6\% of the diffuse flux, respectively, assuming  CSM  interaction.  Assuming the choked-jet scenario, we constrain SNe Ibc to contribute no more than 16.4\%. The low rate of SNe IIn ($\sim$6\% of all SNe) allows for a tight constraint on the flux from the whole population. SNe IIP are the most common subtype (~52\%) and thus the constraint on the population flux is relatively loose. The small time window in the choked-jet scenario excludes more background neutrinos from the arrival time alone which leads to a high sensitivity and a tight constraint on the population's contribution to the diffuse flux. 

\begin{figure}
    \centering
    \includegraphics[width=0.7\textwidth]{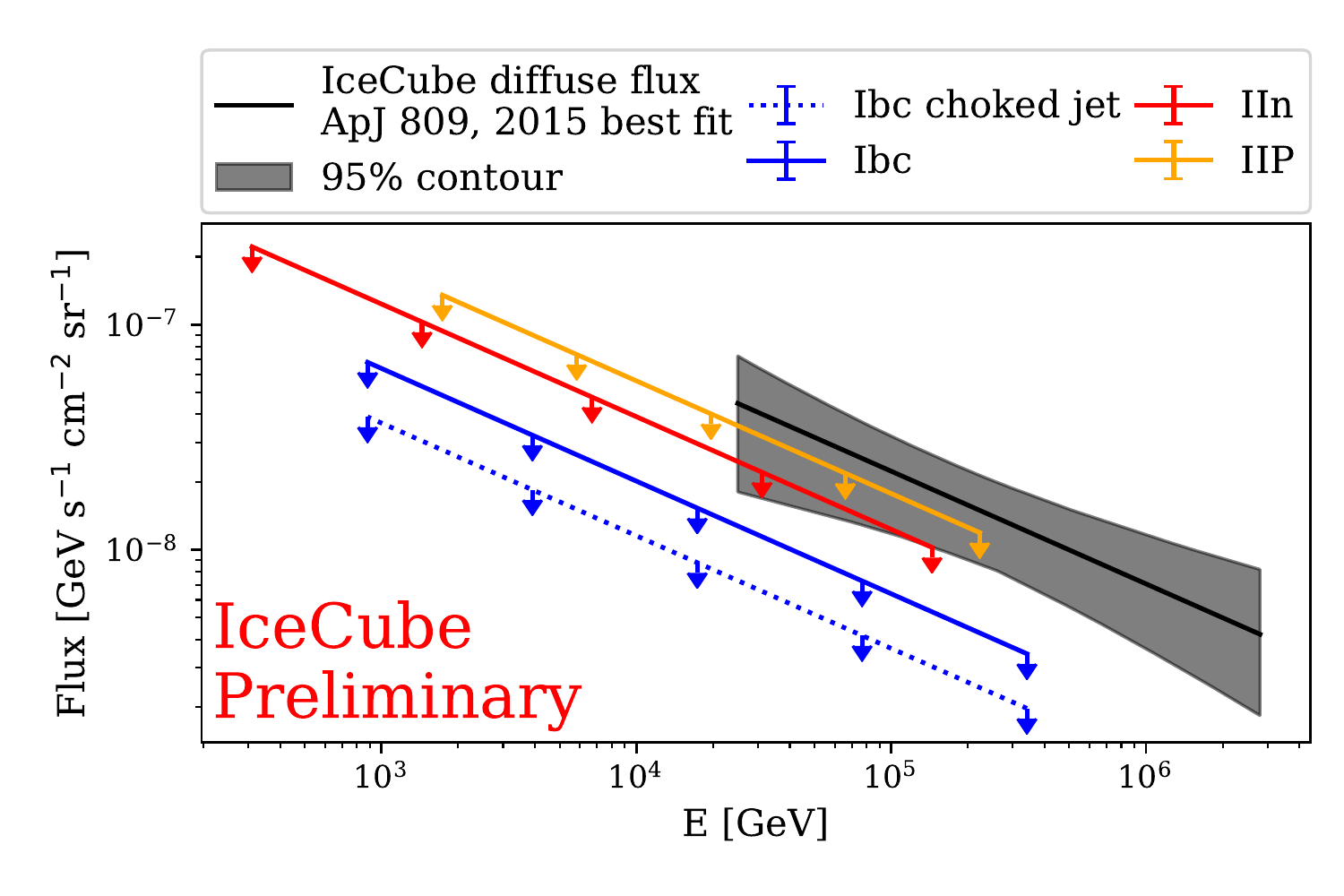}
    \caption{Limits on the contribution to the diffuse flux. Assuming CSM interaction, SNe IIn, IIP and Ibc can only contribute 55.2\%, 79.6\% and 28.6\%, respectively. Assuming the choked-jet scenario, we constrain SNe Ibc to contribute no more than 16.4\%. The diffuse flux measurement is taken from \cite{diffuse_flux_measurement}.}
    \label{fig:flux_limits}
\end{figure}

We calculate the energy range of the analysis by bounding the energy interval from which we select simulated signal events from high and low energies. For both cases we find the energy where the sensitivity drops by 5\%. We define the range between both values as our 90\% sensitive energy region. With that we can derive the limits on the contribution to the diffuse astrophysical neutrino flux as shown in figure \ref{fig:flux_limits}. This limit is of the same order of magnitude as the ~30\% contribution limit from Fermi-detected blazars \cite{fermi_2lac_stacking}.

\bibliographystyle{ICRC}
\bibliography{references}

\clearpage
\section*{Full Author List: IceCube Collaboration}




\scriptsize
\noindent
R. Abbasi$^{17}$,
M. Ackermann$^{59}$,
J. Adams$^{18}$,
J. A. Aguilar$^{12}$,
M. Ahlers$^{22}$,
M. Ahrens$^{50}$,
C. Alispach$^{28}$,
A. A. Alves Jr.$^{31}$,
N. M. Amin$^{42}$,
R. An$^{14}$
% ...
\\

\noindent
% ...
$^{12}$ Universit{\'e} Libre de Bruxelles, Science Faculty CP230, B-1050 Brussels, Belgium \\
% ...
$^{14}$ Department of Physics and Laboratory for Particle Physics and Cosmology, Harvard University, Cambridge, MA 02138, USA \\
% ...
$^{17}$ Department of Physics, Loyola University Chicago, Chicago, IL 60660, USA \\
$^{18}$ Dept. of Physics and Astronomy, University of Canterbury, Private Bag 4800, Christchurch, New Zealand \\
% ...
$^{22}$ Niels Bohr Institute, University of Copenhagen, DK-2100 Copenhagen, Denmark \\
% ...
$^{28}$ D{\'e}partement de physique nucl{\'e}aire et corpusculaire, Universit{\'e} de Gen{\`e}ve, CH-1211 Gen{\`e}ve, Switzerland \\
% ...
$^{31}$ Karlsruhe Institute of Technology, Institute for Astroparticle Physics, D-76021 Karlsruhe, Germany  \\
% ...
$^{42}$ Bartol Research Institute and Dept. of Physics and Astronomy, University of Delaware, Newark, DE 19716, USA \\
% ...
$^{50}$ Oskar Klein Centre and Dept. of Physics, Stockholm University, SE-10691 Stockholm, Sweden \\
% ...
$^{59}$ DESY, D-15738 Zeuthen, Germany \\
% ...
\subsection*{Acknowledgements}

\noindent
USA {\textendash} U.S. National Science Foundation-Office of Polar Programs,
U.S. National Science Foundation-Physics Division,
U.S. National Science Foundation-EPSCoR,
Wisconsin Alumni Research Foundation,
Center for High Throughput Computing (CHTC) at the University of Wisconsin{\textendash}Madison,
Open Science Grid (OSG),
Extreme Science and Engineering Discovery Environment (XSEDE),
Frontera computing project at the Texas Advanced Computing Center,
U.S. Department of Energy-National Energy Research Scientific Computing Center,
Particle astrophysics research computing center at the University of Maryland,
Institute for Cyber-Enabled Research at Michigan State University,
and Astroparticle physics computational facility at Marquette University;
Belgium {\textendash} Funds for Scientific Research (FRS-FNRS and FWO),
FWO Odysseus and Big Science programmes,
and Belgian Federal Science Policy Office (Belspo);
Germany {\textendash} Bundesministerium f{\"u}r Bildung und Forschung (BMBF),
Deutsche Forschungsgemeinschaft (DFG),
Helmholtz Alliance for Astroparticle Physics (HAP),
Initiative and Networking Fund of the Helmholtz Association,
Deutsches Elektronen Synchrotron (DESY),
and High Performance Computing cluster of the RWTH Aachen;
Sweden {\textendash} Swedish Research Council,
Swedish Polar Research Secretariat,
Swedish National Infrastructure for Computing (SNIC),
and Knut and Alice Wallenberg Foundation;
Australia {\textendash} Australian Research Council;
Canada {\textendash} Natural Sciences and Engineering Research Council of Canada,
Calcul Qu{\'e}bec, Compute Ontario, Canada Foundation for Innovation, WestGrid, and Compute Canada;
Denmark {\textendash} Villum Fonden and Carlsberg Foundation;
New Zealand {\textendash} Marsden Fund;
Japan {\textendash} Japan Society for Promotion of Science (JSPS)
and Institute for Global Prominent Research (IGPR) of Chiba University;
Korea {\textendash} National Research Foundation of Korea (NRF);
Switzerland {\textendash} Swiss National Science Foundation (SNSF);
United Kingdom {\textendash} Department of Physics, University of Oxford.

\end{document}